\documentclass[twocolumn,aps,prb]{revtex4-1}
\usepackage{graphicx}
\usepackage{amssymb}
\usepackage{amsmath}
\usepackage{bm}
\usepackage{color}

\def\Journal #1,#2,#3,#4#5#6#7{#1 {\bf #2}, #3 (#4#5#6#7)}

\def\Vec{\mathbf}

\begin{document}

\title{Three-dimensional graphdiyne as a topological nodal-line semimetal \\
}
\author{Takafumi Nomura$^1$, Tetsuro Habe$^1$, Ryota Sakamoto$^2$ and Mikito Koshino$^1$}
\affiliation{$^1$ Department of Physics, Osaka University,  Toyonaka 560-0043, Japan}
\affiliation{$^2$Department of Chemistry, The University of Tokyo, Tokyo 113-0033, Japan}
\date{\today}

\begin{abstract}
We study the electronic band structure of three-dimensional ABC-stacked (rhombohedral) graphdiyne,
which is a new planar carbon allotrope recently fabricated. 
Using the first-principles calculation, we show that the system 
is a nodal-line semimetal, in which the conduction band and valence band cross at a
closed ring in the momentum space.
We derive the minimum tight-binding model 
and the low-energy effective Hamiltonian in a $4\times 4$ matrix form.
The nodal line is protected by a non-trivial winding number,
and it ensures the existence of the topological surface state in a finite-thickness slab.
The Fermi surface of the doped system exhibits a peculiar, self-intersecting 
hourglass structure, which is quite different from the torus or pipe shape in the previously proposed nodal semimetals.
Despite its simple configuration,
three-dimensional  graphdiyne  offers unique electronic properties distinct from any other carbon allotropes.
\end{abstract}



\maketitle



\begin{figure}[h]
\begin{center}
\leavevmode\includegraphics[width=1.\hsize]{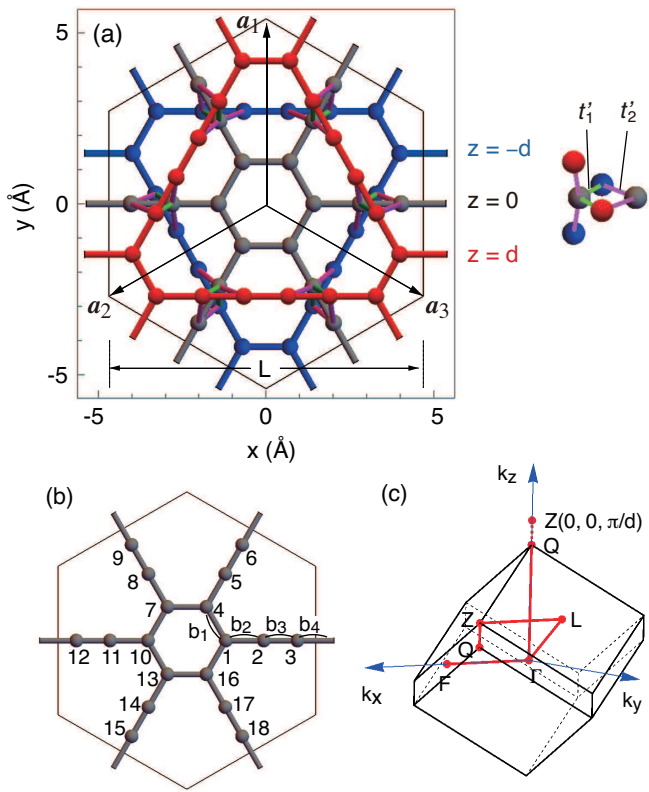}
\end{center}
\caption{ (a) Stacing structure of ABC-stacked GDY projected on $xy$-plane.
(b) A unit cell with numbering of the atoms.
(c) Brillouin zone. The same symbols represent the equivalent point.}
\label{fig_atom}
\end{figure}

Carbon has a wide variety of allotropes 
owing to its ability to form different bonding configurations, $sp$, $sp^2$ and $sp^3$.
Graphene\cite{novoselov2005two} is the most-common two-dimensional carbon  in $sp^2$ bonding,
and a great deal of theoretical and experimental efforts have been devoted 
to explore its unusual physical properties and innovative applications \cite{neto2009electronic, allen2009honeycomb}.
On the other hand, different two-dimensional carbon allotropes, graphyne,
have also been theoretically pursued for decades.
\cite{baughman1987structure,narita1998optimized,narita2000electronic} 
Graphyne is a group of planar carbon systems consisting of benzene rings ($sp^2$ carbons) and ethynyl bridges ($sp$ carbons),
and various geometric structures have been proposed. 
\cite{baughman1987structure,narita1998optimized,narita2000electronic,srinivasu2012graphyne,cranford2012extended,li2014graphdiyne} 
Recently, one of graphyne derivatives, graphdiyne (GDY) \cite{haley1997carbon} [Fig.\ \ref{fig_atom}(b)]
was successfully fabricated by chemical polymerization of organic monomers \cite{li2010architecture},
and so far it is the only graphyne that is realized experimentally. 
The GDY takes a form of three-dimensional stack of monolayers 
\cite{li2010architecture,qian2015self,kuang2015highly, jia2017synthesis, shang2018ultrathin}.
Various applications have been proposed for bulk GDY \cite{kuang2015highly,jia2017synthesis, shang2018ultrathin},
however, its stacking structure was not elucidated. 
Quite recently, one of the authors has produced a high-quality GDY nanosheet, and identified its structure as ABC (rhombohedral) stacking \cite{matsuoka2017crystalline} [Fig.\ \ref{fig_atom}(a)].
In theory, the electronic property was studied for GDY monolayer and multilayers
\cite{narita1998optimized,long2011electronic,luo2011quasiparticle,pan2011graphyne,cranford2012extended,srinivasu2012graphyne,zheng2012structural,li2014graphdiyne,hu2015three,sun2015graphdiyne,jalili2015study},
while the three-dimensional (3D) ABC-stacked GDY has not yet been investigated. 

\begin{figure}[h]
\begin{center}
\leavevmode\includegraphics[width=1.\hsize]{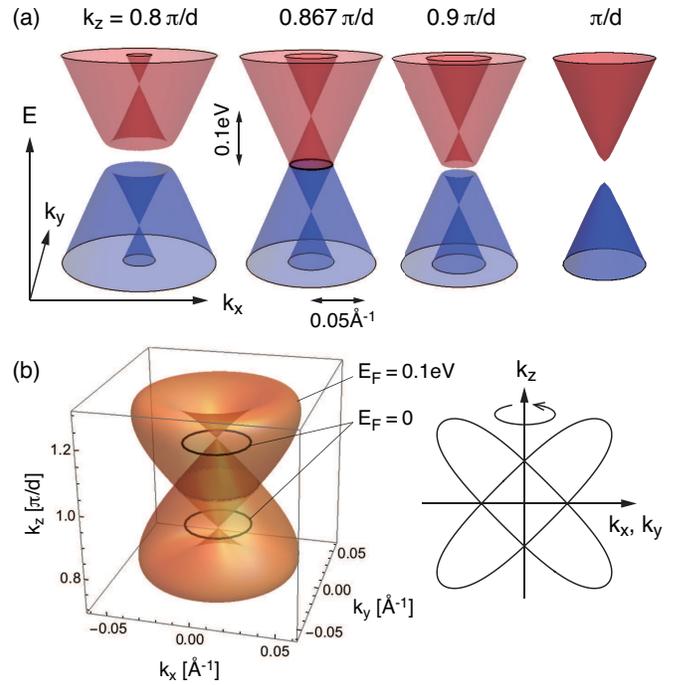}
\end{center}
\caption{(a) Energy band structure of $4\times 4$ effective model for ABC-stacked GDY,
plotted against $(k_x,k_y)$ at various fixed $k_z$'s.
(b) Fermi surface for $E_F=0.1$eV and the nodal lines for $E_F=0$
in the effective model. The right panel is its cross section of the $E_F=0.1$eV surface.
}
\label{fig_surface}
\end{figure}

In this paper, we first study the electronic structure of 3D ABC-stacked GDY
by the first-principles calculation and the effective mass theory. 
In contrast to the semiconducting band structure in monolayer GDY, \cite{narita1998optimized,long2011electronic,luo2011quasiparticle},
3D GDY is found to be a nodal-line semimetal \cite{Burkov2011,Habe2014,Phillips2014,Fang2015},
in which the conduction band and valence band cross at a closed ring in the momentum space
[Fig.\ \ref{fig_surface}(a)].
The nodal-line semimetal is a novel class of topological materials extensively studied in recent years.\cite{Burkov2011,Habe2014,Phillips2014,Fang2015,
PhysRevLett.115.036806,volovik2015standard,chen2015nanostructured,heikkila2015nexus,xie2015new,PhysRevB.93.205132,PhysRevLett.115.026403,koshino2016magnetic,ezawa2016loop,PhysRevB.92.045108,PhysRevLett.116.195501,PhysRevB.93.054520,Schoop2016,Neupane2016} 
We obtain the tight-binding model and the low-energy effective Hamiltonian, with which
we prove the topological protection of the nodal line as well as the existence of the topological surface states.
We also show that the Fermi surface of the doped system exhibits a peculiar self-intersecting hourglass shape
as illustrated in  Fig.\ \ref{fig_surface}(b), which is quite different from torus or pipe shape in the previously proposed nodal semimetals. \cite{Burkov2011,Phillips2014}
GDY is yet another carbon material with a unique electronic structure, which is distinct from any other carbon allotropes.
The discovery of topological nature of GDY would expand the field of topological materials to the diverse class of graphyne derivatives.

The atomic structure of ABC-stacked GDY is illustrated in Fig.\ \ref{fig_atom}(a).
The structure is rhombohedral \cite{matsuoka2017crystalline}
with lattice vectors 
$\Vec{a}_1 = (0,L/\sqrt{3},d)$, $\Vec{a}_2 = (-L/2,-L/(2\sqrt{3}),d)$
and $\Vec{a}_3 = (L/2,-L/(2\sqrt{3}),d)$, 
where $L $ is the lattice constant of single layer GDY
and $d$ is the interlayer spacing. 
A unit cell consists of 18 carbon atoms as shown in Fig.\ \ref{fig_atom}(b), 
where a benzene ring at the center 
is connected to six neighboring cells by linear chains.
As shown later, the distance between neighboring carbon atoms slightly differs
depending on its position, and it is labeled as $b_i\,(i=1,2,3,4)$
as indicated in Fig.\ \ref{fig_atom}(b).


We perform the first-principles density function theory (DFT) calculation 
using the numerical package of quantum-ESPRESSO \cite{Quantum-espresso}
to obtain the electronic structure of three-dimensional ABC-stacked GDY.
Here we employ the ultrasoft pseudopotentials with Perdew-Zunger self-interaction corrected density functional, 
the cutoff energy of the plane-wave basis 60 Ry, and the convergence criterion of 
10$^{-8}$ Ry in $12\times12\times12$ wave number mesh.
In the calculation, the lattice structure and atomic positions 
are optimized by structural relaxation code in quantum-ESPRESSO. 
Here the criterion for the structural relaxation for total energy convergence
is taken as 10$^{-4}$ Ry in $6\times6\times6$ wave number mesh, 
and that for force on atoms is taken as 10$^{-3}$ Ry/$a_B$ with Bohr radius $a_B$.

\begin{figure}[h]
\begin{center}
\leavevmode\includegraphics[width=0.9\hsize]{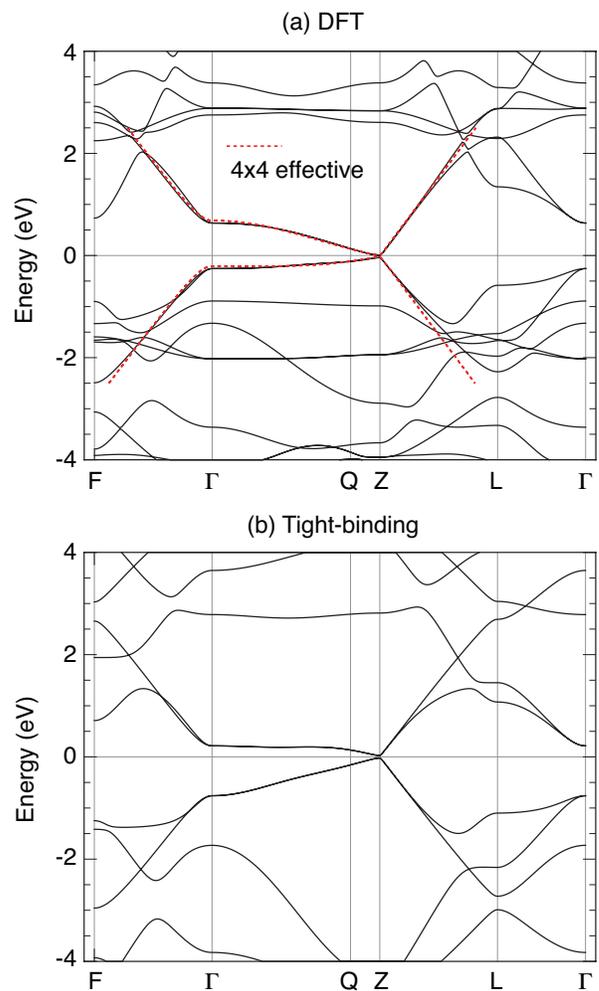}
\end{center}
\caption{(a) DFT-calculated band structure of ABC-stacked GDY 
on the $k$-space path indicated in Fig.\ \ref{fig_atom}(c). 
Red dashed curves are the band of the 4$\times$4 effective model Eq.\ (\ref{eq_eff_ABC}).
(b) Corresponding band structure of the tight-binding model.
}
\label{fig_band_full}
\end{figure}

\begin{figure}[h]
\begin{center}
\leavevmode\includegraphics[width=1.\hsize]{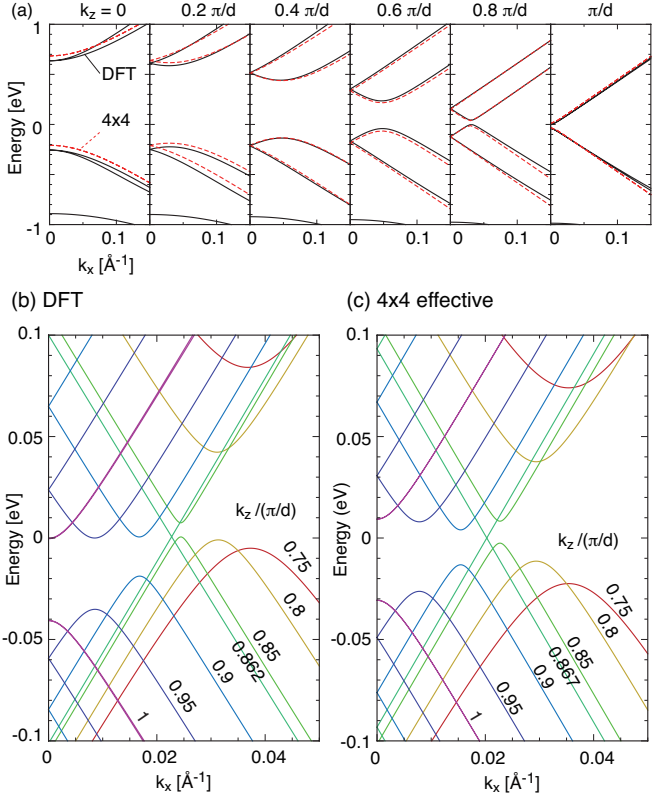}
\end{center}
\caption{
(a) Low-energy band structure of ABC-stacked GDY 
plotted against $k_x$.
Black solid and red dashed curves indicate the DFT and the 4$\times$4 effective model Eq.\ (\ref{eq_eff_ABC}),
 respectively.
(b) Magnified plot near the band closing point for DFT (left) and the effective model (right).
}
\label{fig_band_zoom}
\end{figure}

In the optimized structure, we have $L = 9.38$\AA, $d = 3.29$\AA,
$b_1=1.42$ \AA, $b_2=1.38$\AA, $b_3=1.23$\AA, and $b_4=1.33$\AA,
and there is no out-of-plane distortion.
Figure \ref{fig_band_full}(a) presents the calculated band structure of the ABC-stacked 
GDY on the $k$-space path in Brillouin zone [Fig.\ \ref{fig_atom}(c)]. 
We find that the interlayer coupling significantly reduces the energy gap of monolayer,
and the energy bands are nearly touching at $Z$ point, or $(0,0,\pi/d)$. 
The band crossing is actually located at off-symmetric point near $Z$.
Figure \ref{fig_band_zoom}(a)  shows the band dispersion along the in-plane direction ($k_x$)
at several fixed $k_z$'s from $0$  to $\pi/d$,
and (b) is the further detailed plot near zero energy.
At $k_z=\pi/d$, we have nearly two-fold degenerate bands
in each of the conduction and the valence sectors.
When the $k_z$ shifts away from $\pi/d$, the degeneracy splits
and the middle two bands touch in $k_z \approx 0.862 \pi/d$,
at some off-center in-plane momentum $k_x$.
The low-energy band structure is nearly circular symmetric with respect to $k_z$ axis,
so that the band touching point forms a ring on $k_xk_y$ plane.
The nodal ring is slightly distorted in 120$^\circ$ symmetry,
and also disperses in energy with the width $\sim 1$ meV.

Since the low-energy band structure of GDY is dominated by $p_z$ orbital \cite{narita1998optimized},
it can be described by the single band tight-binding model.
Here we construct a simple tight-binding Hamiltonian to describe ABC-stacked GDY
by considering the major hopping integrals only.
For the intralayer coupling, we take the nearest neighboring hoppings
$t_i\,(i=1,2,3,4)$, which correspond to the distance $b_i\,(i=1,2,3,4)$, respectively.
For the interlayer coupling, we take the two shortest, nearly vertical bonds
of 3.34 \AA and 3.38 \AA,
and define the corresponding hopping integrals as $t'_1$ and $t'_2$, respectively [Fig.\ \ref{fig_atom}(a)].
To obtain the transfer integral of those selected bonds, we adopt the Slater-Koster type formula 
used for carbon $p_z$-orbitals in graphene \cite{slater1954simplified,nakanishi2001conductance,moon2012energy,moon2013opticalabsorption},
 \begin{eqnarray}
&& -t(\Vec{R}) = 
V_{pp\pi}\left[1-(\hat{\Vec{R}}\cdot\Vec{e}_z)^2\right]
+ V_{pp\sigma}(\hat{\Vec{R}}\cdot\Vec{e}_z)^2,
\nonumber \\
&& V_{pp\pi} =  V_{pp\pi}^0 e^{- (R-a_0)/r_0},
\,\, V_{pp\sigma} =  V_{pp\sigma}^0  e^{- (R-d_0)/r_0}.
\label{eq_transfer_integral}
\end{eqnarray}
Here $\Vec{R}$ is the distance between two atoms, $\hat{\Vec{R}}=\Vec{R}/R$ and
$\textbf{e}_z$ is the unit vector on $z$ axis. 
$V_{pp\pi}^0 \approx -2.7$eV is the transfer integrals between nearest-neighbor atoms of monolayer graphene which 
are located at distance $a_0 = a/\sqrt{3} \approx 0.142$nm. $V_{pp\sigma}^0 \approx 0.48 $eV is the transfer integral 
between two nearest-vertically aligned atoms and $d_0 \approx 0.334$nm is the interlayer spacing of graphite. 
The parameter $r_0$, the decay length of transfer integral is chosen as $0.09$ nm.
The hopping parameters (i.e., $t(\Vec{R})$ for selected bonds) are then obtained as
$t_1 = 2.71$ eV, $t_2 = 2.82$ eV, $t_3 = 3.34$ eV, $t_4 = 2.99$ eV, $t'_1 = -0.463$ eV, $t'_2 = -0.421$ eV.
Figure\ \ref{fig_band_full}(b) presents the band structure of the tight-binding model.
By comparing with the DFT band structure in Fig.\ \ref{fig_band_full}(a),
we actually see that the qualitative features of low-energy bands are well reproduced.

Finally we derive the effective continuum Hamiltonian for the low-energy spectrum.
At $\Gamma$-point, we define four bases,
\begin{align}
 |\psi_{1}\rangle =&
(i/\sqrt{18}) (1,1,-1,\omega,\omega,-\omega,\omega^*,\omega^*,-\omega^*\nonumber\\
& \qquad\qquad  1,1,-1,\omega,\omega,-\omega,\omega^*,\omega^*,-\omega^*), \nonumber\\
 |\psi_{2}\rangle =&
(i/\sqrt{18})  (1,1,-1,\omega^*,\omega^*,-\omega^*,\omega,\omega,-\omega,\nonumber\\
&\qquad\qquad  1,1,-1,\omega^*,\omega^*,-\omega^*,\omega,\omega,-\omega), \nonumber\\
 |\psi_{3}\rangle =&
 (1/\sqrt{18})(1,-1,-1,-\omega,\omega,\omega,\omega^*,-\omega^*,-\omega^*,\nonumber\\
& \qquad\qquad  -1,1,1,\omega,-\omega,-\omega,-\omega^*,\omega^*,\omega^*), \nonumber\\
 |\psi_{4}\rangle =&
 (1/\sqrt{18})(1,-1,-1,-\omega^*,\omega^*,\omega^*,\omega,-\omega,-\omega,\nonumber\\
& \qquad\qquad -1,1,1,\omega^*,-\omega^*,-\omega^*,-\omega,\omega,\omega),
\label{eq_ base}
\end{align}
where $\omega = e^{2\pi i/3}$, and the vector components represent the wave amplitudes of the site 1 to 18
in the unit cell [Fig.\ \ref{fig_atom}(b)].
The four bases  are eigenstates of space inversion 
with parity $+$ for $\psi_{1} , \psi_{2}$ and $-$ for $\psi_{3} , \psi_{4}$.
They are also eigenstates of $120^\circ$ rotation with eigenvalue
$\omega$ for $\psi_{1} ,\psi_{3}$ and $\omega^*$ for $\psi_{2} , \psi_{4}$.
The time-reversal operation relates the bases as $\psi_1 = -\psi_2^*$  and $\psi_3 = \psi_4^*$.
The reflection $R_x[(x,y,z)\to(-x,y,z)]$ operates as $R_x\psi_1 = \psi_2$  and $R_x \psi_3 = -\psi_4$.
The bases at general $k$-points are defined by multiplying $\psi_i$ by the Bloch factor
$e^{i\Vec{k}\cdot\Vec{r}}$ for every lattice point $\Vec{r}$.

We set our basis as $U = ( |\psi_{1}\rangle , |\psi_{2}\rangle , |\psi_{3}\rangle , |\psi_{4}\rangle )$
and obtain the reduced $4\times 4$ Hamiltonian by calculating the matrix $U^\dagger H U$ with the tight-binding Hamiltonian $H$. Then we expand the matrix in terms of in-plane wave number $(k_x,k_y)$,
and obtain the effective low-energy Hamiltonian.
For the single layer GDY, the Hamiltonian within the linear order of $(k_x,k_y)$ becomes
\begin{equation}
H_{\rm mono} = 
\begin{pmatrix}
\Delta & 0 & 0& \hbar v k_- \\
0 & \Delta & \hbar v k_+ & 0\\
0 & \hbar v k_- & -\Delta & 0\\
\hbar v k_+ & 0 & 0 & -\Delta  \\
\end{pmatrix}
\end{equation}
where $k_\pm = k_x \pm i k_y$,
$\Delta = (t_1 - 2 t_2 + 2 t_3 - t_4)/3$, 
$v =(2 t_1 + 2 t_2 + 2 t_3 + t_4)b/(6\hbar)$, and
$b$ is the average of neighboring bond lengths $b_1 \cdots b_4$.
It is equivalent with the two-dimensional Dirac Hamiltonian 
with the mass parameter $\Delta$ and the light velocity replaced with $v$.
The mass gap $\Delta$ is sensitive to the 
difference among the interlayer hopping parameters $t_1, \cdots, t_4$;
if they were all the identical, $\Delta$ vanishes and the spectrum would be gapless.



The Hamiltonian of ABC-stacked GDY is derived in a similar manner as
\begin{align}
& H_{\rm ABC} = 
\begin{pmatrix}
\epsilon_0 + m & 0 & i b& \hbar v k_- \\
0 & \epsilon_0 + m & \hbar v k_+ & -i b\\
-i b & \hbar v k_- & \epsilon_0 - m & 0\\
\hbar v k_+ & i b & 0 & \epsilon_0 - m  \\
\end{pmatrix},
\nonumber\\
& \epsilon_0(k_z) = u_0 \cos(k_z d),  \quad m(k_z)= \Delta + u_1 \cos(k_z d), \nonumber \\
& b(k_z)=  u_2 \sin(k_z d).
\label{eq_eff_ABC}
\end{align}
Here $u_0, u_1$ and $u_2$ describe the interlayer coupling,
and they are related by the tight-binding hoppings as
$u_0 = t'_1/3$, $u_1 = -2 t'_2/3$ and $u_2 = - t'_1/\sqrt{3}$.
$\Delta$ and $v$ are just as defined for monolayer.
We can show that the nearest-neighbor interlayer Hamiltonian for the basis of Eq.\ (\ref{eq_ base})
is required to have the above form within zero-th order in $k_x$ and $k_y$, 
only by considering the symmetries mentioned above.
There $u_0$, $u_1$ and $u_2$ are free parameters.

The five parameters $(v, \Delta, u_0, u_1, u_2)$ of $H_{\rm ABC}$
can be determined directly to fit the DFT band structure.
Our best fit is $v \approx 7.0\times 10^5$ m/s, $\Delta = 0.213$ eV,
$u_0 = 0.125$ eV, $u_1 = 0.233$ eV, and $u_2 = 0.230$ eV.
In Fig.\ \ref{fig_band_full}(a) and Fig.\ \ref{fig_band_zoom}(a), 
the red dashed curves indicate the energy bands of the effective band model with those parameters,
which fit quite well with the DFT band structure in the low energy region.
We also show the detailed plot near zero energy in Fig.\ \ref{fig_band_zoom}(c),
where the band touching nature is also reproduced.


The eigenenergies of $H_{\rm ABC}$ is given by
$E_{s,s'}(\Vec{k}) = \epsilon_0 + s \sqrt{m^2+ (\hbar v k_{\parallel} + s' b)^2}$,
where $s=\pm$, $s' =\pm$ and $k_{\parallel} = (k_x+k_y)^{1/2}$.
The energy bands of $s=\pm$ touch
only when $m(k_z)=0$ and $\hbar v k_{\parallel} + s' b(k_z) =0$,
or equivalently, $k_z d = \pm \arccos(-\Delta/u_1)$ and $\hbar v k_{\parallel} =  u_2 \sqrt{1-\Delta^2/u_1^2}$.
This determines the position of the nodal line.
The real solution for $(k_\parallel, k_z)$ exist when $|\Delta / u_1| \leq 1$,
which is met in our system. 
The topological protection of the nodal line is checked by calculating the Berry phase
along $k_z$ axis at fixed in-plane momentum $(k_x,k_y)$.
We define a unitary matrix,
\begin{align}
V = 
\frac{1}{\sqrt{2}}
\begin{pmatrix}
1 & 0 & 1 & 0 \\
-i e^{i\theta} & 0 & i e^{i\theta} & 0 \\
0 & 1 & 0 & 1 \\
0 & i e^{i\theta} & 0 & -i e^{i\theta} \\
\end{pmatrix},
\end{align}
where $\tan \theta = k_y/k_x$.
Then the unitary transformed Hamiltonian $V^\dagger H_{\rm ABC} V$ is 
block-diagonalized into two $2\times2$ matrices $H_\pm = Y\sigma_y + Z \sigma_z$, 
where $Y = \pm \hbar vk_\parallel - b(k_z)$ and $Z = m(k_z)$,
and $\sigma_y$ and $\sigma_z$ are the Pauli matrices.
When $k_z$ is changed from $-\pi/d$ to $\pi/d$, the trace of $(Y,Z)$
encloses the origin only when $|\Delta / u_1| \leq 1$ and $\hbar v k_{\parallel} < u_2 \sqrt{1-\Delta^2/u_1^2}$,
so that the Berry phase is $\pi$ only when the integral path passes through the nodal ring
while it is 0 otherwise.
Since the system has the time-reversal symmetry and the space inversion symmetry,
the Berry curvature always vanishes at any nondegenerate points in the energy band \cite{haldane2004berry,fu2007topological}, 
and this guarantees the robustness of band touching
against a small perturbation not to break these symmteries \cite{koshino2013electronic}.

\begin{figure}[h]
\begin{center}
\leavevmode\includegraphics[width=0.7\hsize]{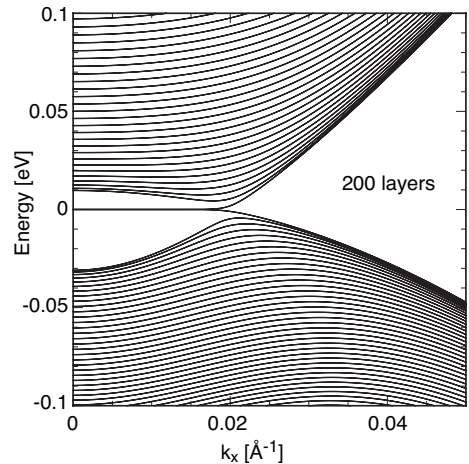}
\end{center}
\caption{Energy bands of ABC-stacked GDY with thickness of 200 layers}
\label{fig_slab}
\end{figure}

The nodal line also implies the existence of the topological surface states.
Figure \ref{fig_slab} shows the band structure calculated for the effective Hamiltonian with 
a finite thickness of 200 layers, plotted against $k_x$ axis.
We actually see that nearly-flat band originated from the surface modes
appear between the nodal points. In $(k_x, k_y)$-space, 
the surface-state band spans the disk-shaped region inside the nodal ring.


When the Fermi energy is shifted from the nodal line by doping electrons or holes, 
the Fermi surface with a nontrivial structure emerges.
Figure \ref{fig_surface}(b) shows the Fermi surface 
for $E_F=0.1$ eV in the effective model.
The figure is rotationally symmetric with respect to $k_z$ axis,
and its vertical cross section is presented in the right panel.
The surface has a pair of conical points on $k_z$ axis,
at which the electron-like pocket and the hole-like pocket are connected.
The surface is self-intersecting at $k_z = \pi/d$, 
and this corresponds to the degeneracy of the two conduction bands
at $k_z = \pi/d$. In the higher Fermi energy, the energy bands and Fermi surface are
trigonally warped in 120$^\circ$ symmetry,
while the warping is irrelevant in $|E_F| < 0.1$eV and neglected in the present effective model.
We expect that the peculiar self-intersecting Fermi surface 
would cause unusual effects on the magnetotransport and optical properties through 
the formation of the self-connected semiclassical orbits \cite{o2016magnetic, koshino2016cyclotron,bovenzi2018twisted}.

The spin-orbit coupling is entirely neglected in the present calculation. 
It tends to gap out the nodal line in the absence of non-symmorphic symmetry \cite{Fang2015}.
In carbon materials, however, the spin-orbit coupling is relatively small, 
and in particular, it is extremely weak in flat systems where $\pi$-band and $\sigma$-band are 
independent from each other \cite{huertas2006spin}.
For example, the spin-orbit energy scale is only 10 mK in graphene \cite{huertas2006spin,min2006intrinsic}.
A similar discussion may be applicable to GDY, a flat carbon $\pi$-system, and therefore 
the predicted nodal line should be preserved. 


To conclude, we calculated the electronic structure of three-dimensional GDY, and found it to be 
a topological nodal-line semimetal with unique self-intersecting hourglass Fermi surface.
We derive the minimal effective low-energy Hamiltonian 
and proved the topological protection of nodal line and also existence of the topological surface states.
The graphyne family has an enormous variety of derivatives with various geometric structures and atomic species \cite{baughman1987structure,srinivasu2012graphyne,cranford2012extended,li2014graphdiyne}.
The discovery of topological nature of GDY would expand the diversity of topological matters 
by offering a vast unexplored field.


M. K. and T. H. acknowledge support of JSPS KAKENHI Grant Numbers
JP25107005, JP25107001 and JP17K05496.
R. S. acknowledges support of JSPS KAKENHI Grant Numbers JP16H00900 and JP26708005,
JST PRESTO Grant Number JPMJPR1516, Japan,
the Asahi Glass Foundation, Kato foundation for Promotion of Science, the Murata Science Foundation,
Yashima Environment Technology Foundation and Foundation Advanced Technology Institute.

\bibliography{graphdiyne}

\end{document}